\documentclass[aip,reprint]{revtex4-1}
\usepackage{amsfonts}
\usepackage{amsmath,amssymb}
\usepackage{graphicx}

\draft 

\begin{document}


\title{Dichotomous noise models of gene switches} 



\author{Davit A Potoyan}
\email[]{potoyan@rice.edu}
\affiliation{Department of Chemistry and Center for Theoretical Biological Physics, Rice university, Houston TX 77005 }

\author{Peter G Wolynes}
\email[]{pwolynes@rice.edu}
\affiliation{Department of Chemistry and Center for Theoretical Biological Physics, Rice university, Houston TX 77005 }


\date{\today}

\begin{abstract}

Molecular noise in gene regulatory networks has two intrinsic components, one part being due to fluctuations caused by the birth and death of protein or mRNA molecules which are often present in small numbers and the other part arising from gene state switching, a single molecule event.  Stochastic dynamics of gene regulatory circuits appears to be largely responsible for bifurcations into a set of multi-attractor states that encode different cell phenotypes. The interplay of dichotomous single molecule gene noise with the nonlinear architecture of genetic networks generates rich and complex phenomena. In this paper we elaborate on an approximate framework that leads to simple hybrid multi-scale schemes well suited for the quantitative exploration of the steady state properties of large-scale cellular genetic circuits.Through a path sum based analysis of trajectory statistics we elucidate the connection of these hybrid schemes to the underlying master equation and provide a rigorous justification for using dichotomous noise based models to study genetic networks. Numerical simulations of circuit models reveal that the contribution of the  genetic noise of single molecule origin to the total noise is significant for a wide range of kinetic regimes. 
\end{abstract}

\pacs{}

\maketitle 

\section{Introduction}

Molecular noise functions as a driving force for phenotypic heterogeneity in cell populations~\cite{kaern2005stochasticity, eldar2010functional,huang2009non, larson2009single}. With the growth of sophisticated imaging methods and biochemical techniques that probe gene expression at a single cell level, the omnipresence of molecular noise in cellular processes has become widely appreciated~\cite{kaern2005stochasticity,sanchez2013regulation}. Noise in biology therefore  should no longer be viewed as a nuisance with which cells must cope but rather it must be viewed as a natural facilitator for adaptation, growth and development~\cite{eldar2010functional,acar2008stochastic,maheshri2007living,macneil2011gene}. Molecular noise in genetic networks has at least two sources~\cite{paulsson2004summing,graslund2010single}: one source is the probabilistic nature of molecular associations  such as the binding/unbinding of promoters to operator sites on the DNA while the other obvious source is randomness in the sequential reactive events such as the stochastic synthesis and degradation of mRNA, protein molecules, which generate the whole population of molecules. The dichotomous Markov noise (DMN) that arises from a single copy of a gene stochastically switching between its $ON$ and $OFF$ states, whether through promoter binding or through architectural changes of chromatin, is of a qualitatively different mathematical character from the other sources of molecular noise which may be treated as controllably small perturbations when the molecular populations of mRNA and proteins are sufficiently large~\cite{kepler2001stochasticity}. 

Dichotomous gene noise leads to bursty intermittent production of mRNAs. The intermittency arises because mRNA molecules are rapidly synthesized in large groups when the gene becomes activated  but these bursts are then followed by intervals of silence when the gene is turned off. "Poissonian" stochastic kinetics arises when the mRNA's are synthesized independently from each other at random intervals with uniform probability. In contrast, single cell experiments provide ample evidence that the statistics of gene expression is often non-Poisonian, with bursty behavior having been observed in both prokaryotic~\cite{golding2005real,chong2014mechanism,levens2014new,yu2006probing} and eukaryotic~\cite{raser2004control,raj2006stochastic,suter2011mammalian,bothma2014dynamic,larson2011real} cells. As a rule the bursty stochastic kinetics leads to higher overall levels of noise. The detailed microscopic origins of dichotomous gene noise and the observed transcriptional and translational bursts still remain under debate~\cite{chong2014mechanism}. Possible mechanisms include local chromatin remodeling, thermally induced nucleosome turnover, covalent modifications and  genome wide chromatin fiber dynamics~\cite{coulon2013eukaryotic}. 

Dichotomous noise generally breaks detailed balance in the circuits and thus creates  non-equilibrium steady states which can not always be described by quasi-equilibrium fluctuation statistics. Dichotomous noise driven phenomena include robust phase synchronization~\cite{roussel2001phase,yamada2006stochastic}, stochastic hypersensitivity~\cite{tarlie1998optimal,PhysRevLett.80.4840}, enhanced stochastic resonance~\cite{PhysRevE.62.R3031}, hysteresis~\cite{PhysRevE.69.061106} and patterning~\cite{PhysRevE.87.062924,buceta2003spatial}.  

Brute force simulation of the full master equation for genetic networks has already yielded many insights~\cite{gillespie2007stochastic,mcadams1997stochastic}. Nevertheless in practice the sheer number of cellular components often renders such an explicit approach overwhelming. Most conventional approximations for treating stochastic chemical reactions, such as size expansion methods~\cite{elf2003fast,van2011stochastic} are not applicable when there is sufficient dichotomous noise since these approximations usually rely on the uniform scaling  with size of the entire population of species to the near deterministic limit~\cite{Potoyan2014dephasing}. Time averaging can play a role however in reducing the influence of the single molecule dichotomous noise. In the limit of infinitely fast switching of the gene one can again obtain the strictly deterministic result if the molecular populations of all the other components are sufficiently large. This fast gene switching regime has come to be known as the adiabatic limit~\cite{sasai2003stochastic}. In the adiabatic limit there is a very large time-scale separation between gene switching at the single molecule level and the biochemical synthesis and degradation reactions in the rest of the network which can then again be treated as nearly deterministic in the large system size limit.  Thus it is most important to develop approximations appropriate to the non-adiabatic case. 

Recently several schemes for modeling gene networks have been proposed which offer various ways of treating the stochastic aspects of gene switching~\cite{newby2014bistable,newby2014asymptotic,radulescu2012reduction,zhang2014stem, lu2014construction, assaf2011determining,aurell2002epigenetics,PhysRevLett.114.078101}. Most of these schemes still however rely on there being a sufficient degree of time scale separation between gene and protein degrees of freedom, which means that switching dynamics must usually be assumed to be very fast. While such an assumption could be a sound one for many gene circuits, the intermediate regime where there is no such time scale separation is also of interest especially for understanding the behavior of mammalian gene circuits. In this contribution we present a framework which lifts the requirement for the genes to operate near the adiabatic limit. This framework leads to a hybrid multi-scale scheme for numerically exploring potentially large scale genetic networks. We demonstrate the usefulness of our scheme by analyzing some well known models of simple genetic circuits.

\section{Results}
\subsection*{Schematic discussion of trajectory statistics in gene networks.}

In this section we discuss the notion of trajectory probabilities for stochastic gene circuits with dichotomous elements. This notion motivates hybrid simulation schemes based on  piecewise deterministic systems of equations.  We illustrate the ideas using a simple one component model of a self regulating gene as the prototype for a gene network.  Due to its simplicity this idealized problem has been the subject of numerous theoretical studies~\cite{hornos2005self,grima2012steady,walczak2005absolute, kepler2001stochasticity,shi2011perturbation, feng2010adiabatic,kumar2014exact}. The state of the circuit is defined by two dynamical variables: an operator state variable $s=\left\{0,1\right\}$ describing whether the gene is active $s=1$ (ON) or inactive $s=0$ (OFF) and the number of proteins $n=\left\{0,1,2... \infty \right\}$. Since the gene circuit is stochastic, one is interested among other things in knowing the probability $p_s(n,t)$ of having $n$ proteins and the gene being in state $s$ at time $t$. At this level the most satisfying description of such a circuit (if one assumes a well stirred mixture of molecules) is given by a discrete space master equation: 
\begin{widetext}
\begin{eqnarray}
 \frac{\partial P(n,t) }{\partial t} = \mathbb{G}(P(n-1,t)-P(n,t))+\mathbb{K}((n+1)P(n+1,t)-nP(n,t))+\mathbb{W}_n P(n,t)
 \label{mast}
\end{eqnarray}
\end{widetext}
where $P(n,t) =|p_{0}(n,t), p_{1}(n,t)\rangle $ is the state vector specifying the probabilities for each gene/protein state. The diagonal birth/death matrices $\mathbb{G}=diag(g_0,g_1)$ and $\mathbb{K}=diag(k,k)$  contain the rates of synthesis $g$ and degradation $k$ which are assumed to be Poisson processes, while $\mathbb{W}_n$  is the rate matrix that describes gene state switching probabilities as a function of protein numbers. In the case when the populations of the species are not too small ($n\gg 1$) such a discrete representation defined with integer population number accuracy(Eq.~\ref{mast})  may no longer  be absolutely necessary.  When the populations of species in the genetic network are sufficiently large one may instead treat $n$ as a continuous variable and use diffusive dynamics for the evolution of protein population. This approximation is well justified by size expansion arguments. Numerical simulations~\cite{grima2011accurate} show that Fokker-Planck approximations can be quantitatively adequate even when protein numbers are as small as $100$. In this limit  one may reduce the fully discrete master equation description to a master equation in mixed continuous and discrete variables  $\frac{\partial P(n,t) }{\partial t} = -\nabla {J}+\mathbb{W}_n P(n,t)$, where the birth/death terms are replaced with $J=-A(n)-\frac{1}{2}\partial_n D(n)$ corresponding to an effective flux of continuous number of molecules To visualize  trajectories generated by such dynamics consider a particular realization of the gene circuit's history that goes through a particular sequence of gene state flips: $\sigma=\{s_0 \rightarrow s_1 \rightarrow s_2 \rightarrow  ... \rightarrow s_N\} $. These are governed by the master equation for the state variable conditioned at any time by the species population$\dot p_{s'}(n,t')  =\mathbb{W} p_s(n,t)$. For the present model this conditional master equation reads explicitly:
\begin{equation}
\begin{pmatrix} \dot p_0(n,t) \\ \dot p_1(n,t) \end{pmatrix} = 
  \begin{matrix}\\\mbox{}\end{matrix}
  \begin{pmatrix} -k_{off} & k_{on}(n) \\ k_{off} & -k_{on}(n) \end{pmatrix} 
  \begin{pmatrix}  p_0(n,t) \\ p_1(n,t) \end{pmatrix}
  \label{matrix1}
\end{equation}
 In between gene flipping events the protein population will evolve diffusively $\nu=\{n_0 \rightarrow n_1 \rightarrow n_2 \rightarrow  ... \rightarrow n_N\} $ according to one of the following Fokker-Planck equations:  
\begin{equation}
\begin{array}{rcr} 
\partial_t p_0(n,t) & = & \big [ -\partial_{n} A_0(n) + \frac{1}{2}\partial^2_n D_0(n)\big] p_0(n,t)\\ 
\\
\partial_t p_1(n,t) & = & \big [ -\partial_{n} A_1(n) + \frac{1}{2}\partial^2_n D_1(n)\big] p_1(n,t) \end{array}
\label{fokkers}
\end{equation}
where the $A_s(n)$ and $D_s(n)$ stand for the drift and the diffusion coefficients that correspond to the gene state $s$ at any time. For the self regulated gene model these terms are $A_s=g_s-kn$ and $D_s(n)=g_s+kn$. In this example $g_s$ is the protein synthesis rate when the gene is in state $s$ while $k$ is the rate coefficient corresponding to a first order model of protein degradation~\cite{walczak2005absolute,kepler2001stochasticity}. 
Equations 1-2 completely define the dynamics of the simplest gene circuit. The system of equations 1 and 2 is known in mathematics as a coupled or switched diffusion process~\cite{hara1982model,min1984multi,qian2000mathematical}. It features prominently in the problems of molecular motors, ratchets and gene networks. The coupled diffusion process is  analytically intractable for all but the simplest systems. 

Now let us consider a particular realization of the stochastic trajectory of the gene state and then ask what is the probability given this  gene state trajectory of observing a certain sequence of corresponding protein numbers $(n_1, n_2, .. n_N)$ at  times
$(t_1,t_2,...t_N)$. Most experimental studies measure time dependence of protein concentrations rather than follow the gene state changes explicitly. First let us consider a system with a fixed gene state (e.g let's say $s=1$ at all times). Using a convenient bra ket notation the conditional probability that protein state $n_{i+1}$ follows $n_{i}$ at an earlier time can be expressed as $p(n_{i+1}, t_{i+1} | n_i, t_i) =\langle n_{i+1} | {\hat U}(\Delta t) | n_{i} \rangle$  where ${\hat U}(\Delta t)$ is the evolution operator of the Fokker-Planck equation corresponding to the particular fixed gene state. If we then allow the gene state to change  in between gene switching events the diffusion of protein numbers $n_0 \xrightarrow{s_0} n_1 \xrightarrow{s_1}  ...$ will be governed by one of the Fokker Planck equations (Eq \ref{fokkers}). The  evolution operator ${\hat U_{s}}(\Delta t)$ will therefore  depend on the gene states $s$ at each time $t$ as indicated by its subscript.  Thus for any particular sequence of gene states $\sigma$ we have propagators $p_{s_i}(n_{i+1} t_{i+1} | n_{i}, t_i) =\langle n_{i+1} | {\hat U}_{s_i} | n_{i} \rangle $ defined on the intervals between gene switching events. Similarly for a particular realization of protein numbers $\nu$ at times $t_i$ we can write the probabilities of the gene switching, $p_{n_i}(s_{i+1} t_{i+1} | s_{i}, t_i) =\langle s_{i+1} | G_{n_{i}}  | s_{i} \rangle$. Again this is defined on the intervals between protein number diffusion events. Suppose we are given set of protein numbers observed at regular time intervals $\Delta t$ , $\nu=\{n_0,  n_1, n_2,  ... n_N\} $ which could be obtained by recording a stochastic trajectory from a numerical simulation on a computer or by observing the gene expression profile in a single cell in the laboratory. Using the assumed Markovian property of the underlying molecular stochastic processes  we can write the probability of observing a trajectory $\nu$  in terms of the protein states irrespective of the gene trajectory as the product of transition probabilities between protein and gene states, which is then summed over all possible realizations of the gene state trajectory $\sigma=\{s_0, s_1 , s_2 , ..., s_N\} $. 
\begin{widetext}
\begin{equation}
p(n, n_{i-1}, ..., n_0)=  \sum_{\sigma} \prod\limits_{i=0}^{N-1}  \langle s_{i+1} | G_{n_{i}}  | s_{i} \rangle \langle n_{i+1} | U_{s_i} | n_{i} \rangle p(s_0,n_0)
\label{path}
\end{equation}
\end{widetext}
The evolution operators for gene and protein states are:  $G_{n_{i}} = e^{-\mathbb{W}_{n_{i}}\Delta t_i}$ and $U_{s_{i}} = e^{-{\hat L_{s_i}}\Delta t_i}$  with $\mathbb{W}_{n_i}$ being the rate matrix for gene states transitions at  fixed protein state $n_i$ while ${\hat L_{s_i}}$ is the Fokker-Planck operator or protein diffusion corresponding to a fixed gene state $s_i$ for times $t_i$.  Explicitly the short time protein state propagator can be approximated by an Onsager-Machlup like Lagrangian function: 
\begin{widetext}
\begin{equation}
\langle n_{i+1} | e^{-{\hat L_{s_i}}\Delta t_i} | n_{i} \rangle = \mathcal{N}(\Delta t_i) exp{\Big[-\frac{1}{2 D_{s_i}(n_i)} \Big (\frac{n_{i+1}-n_i}{\Delta t_i}-A_{s_i}(n_i)\Big)^2 \Delta t_i \Big]} dn_i
\end{equation}
\end{widetext}
The term  $\mathcal{N}(\Delta t_i)$ takes care of normalization and we have omitted the Jacobian and gauge terms in the Lagrangian that come from the nonlinearity of variable  transformations and the multiplicative nature of noise. These corrections  are vanishingly small in the small noise limit which we are going to take later~\cite{tang2014summing, aurell2002epigenetics}. The terms $A_{s_i}$ and $D_{s_i}$ stand for drift and diffusion coefficients corresponding to gene state $s_i$ at time $t_i$. We may formally take the limit of $\Delta t_i \rightarrow 0$ and write the transition probabilities between protein numbers in a  path integral form:
\begin{widetext}
\begin{equation}
p(n, n_{i-1}, ..., n_0)  = \sum_{\sigma} \prod\limits_{i=0}^{N-1} \langle s_{i+1} | e^{-{\mathbb{W}_{n_i}}\Delta t_i} | s_{i} \rangle e^{-\Delta t_i \mathcal{L}_{s_i}(n_i, n_{i+1}) } = \Big\langle  e^{\int^{n}_{n_0} dt  \mathcal{L}_{s}(\dot n, n, t)} \Big\rangle_{\sigma}
\label{path_int}
\end{equation}
\end{widetext}
We consider the full trajectory probabilities rather than two time probabilities because after averaging  over the gene states in Eq~(\ref{path}) the reduced probability $p(n, n_{i-1}, ..., n_0)$ no longer obeys Markovian rules when the dichotomous noise has a finite correlation time. 
The eigenvalues $\lambda_m (n_{i})$ of the gene state change rate matrix $\mathbb{W}_{n_{i}}$ are $ \lambda_0=0$ which corresponds to a steady state and $\lambda_1=-(k_{on}+k_{off})$, the next largest eigenvalue which sets the time-scale $\tau_d=\lambda_1^{-1}$ for gene state equilibration. The limit of $\tau_d \rightarrow 0$ corresponds to the adiabatic regime while $\tau_d \rightarrow \infty$ corresponds to the non-adiabatic regime of gene expression. 
\begin{figure}[!h]
\includegraphics[width=.45\textwidth]{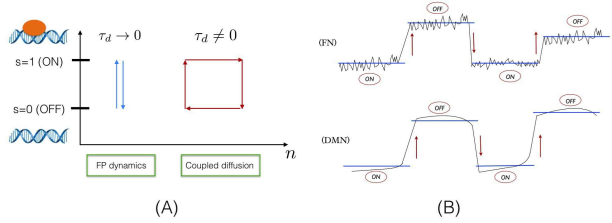}
\caption{ (A) Pictorial view of how dichotomous noise generates churning cycles in the regime with finite gene switching $\tau_d$ (B) Top figure  shows the dynamics with full noise (FN) generated by a master equation which accounts for diffusive motion of protein population and  discrete jump transitions of the gene. In the bottom figure (DMN) the diffusive dynamics is replaced by deterministic equations of motion, while gene transitions are generated from the master equation while coupled to its deterministic counterpart.}
\label{fig:cycle}
\end{figure}
To investigate how the timescale $\tau_d$ of the dichotomous noise affects the trajectory probabilities we can perform a cumulant expansion of the exponential average (\ref{path_int}). 
 \begin{equation} 
\Big\langle e^{\int^{n}_{n_0} dt \mathcal{L}_{s}(\dot n, n, t)} \Big\rangle_{\sigma}={ exp \Bigg({\Big\langle \int^{n}_{n_0} dt \mathcal{L}_{s}(\dot n, n, t)\Big\rangle_{\sigma}+\sum\limits^{k=\infty}_{k=2} \frac{C_k}{k!}}} \Bigg)
\label{cumul}
\end{equation}
We can see that in the adiabatic limit of  fast gene switching ($\tau_d\to0$) the higher order cumulants will vanish because the state of the gene is instantaneously equilibrated at each time.  In this regime the trajectory probability $p(n, n_{i-1}, ..., n_0) $ can be approximated by the first cumulant term which corresponds to gene state averaged stochastic dynamics of protein numbers. Put another way the gene states cease to be correlated at subsequent time intervals $\Delta t$ hence the probability $p(n, n_{i-1}, ..., n_0) $ acquires a markovian property and is therefore reduced to an effective Fokker-Planck description (Fig. \ref{fig:cycle} ). The correlation function of the dichotomous gene variable  ~\cite{horsthemke1984noise} is given by $C_{DMN}(\Delta t) =\frac{D}{\tau_d} exp(-\frac{\Delta t}{\tau_d})$, where $\tau_d=\frac{1}{k_{on}+k_{ff}}$ is the gene switching timescale and $D=k_{on} k_{off} \tau^{3}_d (g_{on}-g_{off})^2$ is the strength of the dichotomous noise. The explicit expression for $D$ in the context of stochastic gene expression was earlier obtained by~\cite{walczak2005absolute} where the term "churning noise" was coined, emphasizing that cycles in the space of gene state and protein number variables generated by alternating  seemingly futile processes of binding/unbinding and protein production/degradation but that nevertheless these cycles leads to dispersion in the probability space. In the case of very slow gene switching ($\tau_d\to\infty$) the memory of a gene state is preserved for a long time thus the trajectory probability can be expressed by a sum of two terms $\Big\langle e^{\int^{n(t)}_{n_0(t_0)} dt \mathcal{L}_{s}(\dot n, n, t)} \Big\rangle_{\sigma} \approx p_{0}e^{\int^{n(t)}_{n_0(t_0)} dt \mathcal{L}_{0}} + p_{1}e^{\int^{n(t)}_{n_0(t_0)} dt \mathcal{L}_{1}}$ corresponding to diffusive motion of protein numbers with the initial state of the gene being ieither $OFF$ or $ON$ with probabilities $p_0$ and $p_1$ respectively. In this extreme non-adiabatic regime one can simply model the dynamics of the gene circuit via Fokker Planck equations for each gene state independently. 

In the intermediate regimes (with finite $\tau_d$), however, a significant component of a gene circuit's stochasticity comes from cyclic motion in the protein number space and one needs to consider contributions from trajectories that perform such driven cyclic drift (Fig. \ref{fig:cycle} ). The contribution of cyclic motion is contained in the higher order cumulants.  We now outline a particularly simple recipe for numerically sampling stochastic trajectories in the intermediate regime of gene switching which accounts for the cyclic motion in the protein number space.     

So far, other than the diffusive assumption for  species represented by continuous population, no major approximations have been invoked in deriving the expression for the short time propagator for the coupled white-dichotomous noise dynamics of genetic circuit. We can further approximate of gene path sum by taking the limit of small birth/death noise $D(n)\rightarrow 0$, which removes the protein fluctuations while preserving only the cyclic component of motion (Fig 1). In this limit the path integrals for each realization of gene state trajectory $\sigma$ in Eq~\ref{path_int} are replaced with deterministic trajectories and the whole path sum is dominated by the stochastic trajectories which follow piecewise jump dynamics in the space of protein numbers:
\begin{subequations}
\begin{align}
        \frac{d n}{dt} &= A(s, n)\\
       \frac{dp_n(s,t)}{dt} &= {\mathbb{W}_n} p_n(s,t)
\end{align}
\label{dmn}
\end{subequations}
Eq~\ref{dmn}a which governs protein dynamics completely ignores fluctuations in protein number that arise from birth/death processes. One may account for birth/death noise by adding a fluctuating linear term correction as noise $B_s \eta(t)$ that stems from the size expansion approximation~\cite{elf2003fast} to each deterministic mode $A_s$. 
\begin{subequations}
\begin{align}
        \frac{d n}{dt} &= A(s, n)+B_s \eta(t)\\
       \frac{dp_n(s,t)}{dt} &= {\mathbb{W}_n }p_n(s,t) \\ 
       \langle \eta(t)\eta(t') \rangle & =\delta(t-t')
\end{align}
\label{dmn_lna}
\end{subequations}
The system of equations \ref{dmn_lna} is one of the key results of this work focusing on gene networks. We note that piecewise deterministic equations and other hybrid equations similar to Eq.~\ref{dmn} and Eq.~\ref{dmn_lna} have already been used in the study of stochastic ion channels, neuronal dynamics, chemical reactions and finance~\cite{kang2013separation, crudu2009hybrid}.  An interesting application of a hybrid scheme to model stochastic bursting effects in toggle switches can be found in the work by Bokes et al~\cite{bokes2013transcriptional} . Particularly attractive hybrid schemes for studying general reaction systems can be found in the works of Salis-Kaznessis~\cite{salis2005accurate} and Haseltine-Rawlings~\cite{haseltine2005origins}. In those schemes the master equation of the entire system is partitioned into fast and slow reaction blocks which are respectively modeled with a jump and continuous state Markov processes. Moreover, the partitioning of the reactions is dynamically adjusted over the course of simulations in order to achieve high quantitative accuracy.  These and many other hybrid schemes while differing in the algorithmic details have the same underlying theme of exploiting time scale separation present in reactions for constructing physically and numerically sounds approximations. The distinguishing features of the dichotomous noise based models proposed in this work are the emphasis on  conceptual simplicity and the gene centric treatment of reactions where dichotomous noise is used exclusively for reactions involving gene states regardless of the rates of other reactions and white noise is used to model fluctuations in the molecular populations of all other species. Thus the proposed scheme is specifically aimed at genetic networks as a relatively coarse-grained (compared to kinetic Monte Carlo and other stochastic schemes) yet fairly simple explorative tool for studying steady state attractors and the associated intrinsic noise caused by the discrete dynamics of the promoter architecture (see the section below for the numerical algorithm). In the context of genetic networks we would like to single out the related works by Ge et al~\cite{PhysRevLett.114.078101} who recently derived a fluctuation model for the self regulating gene model to study the kinetics of barrier crossing in the intermediate regime of gene switching. Their fluctuation model is a particular application of Eq~\ref{dmn}. It is also worth mentioning the works by Karmakar-Bose~\cite{karmakar2004graded} and Zeiser et al~\cite{zeiser2009hybrid} who derived closed analytic expressions for moments and steady state distributions for hybrid models of self regulating gene. 
 In the subsequent sections we will be referring to the sets of equations \ref{dmn} and \ref{dmn_lna} as the dichotomous Markov noise (DMN) and the dichotomous + linear noise  (DMN+LNA) approximations respectively.  Simulating DMN or DMN+LNA schemes can offer a computationally efficient alternative to  direct kinetic MonteCarlo simulations of the full master equation. Lastly the noise obtained via either the DMN or the DMN+LNA scheme can be used in order to decompose the total noise into genetic and non-genetic contributions via $\sigma_{tot}^{2}=\sigma^{2}_{DMN}+\sigma^{2}_{BD}$, where $\sigma^{2}_{DMN}$ is the variance that can be generated by the DMN scheme, $\sigma_{tot}^{2}$ is the variance computed at the level of master equation and $\sigma^{2}_{BD}$ is the non-genetic noise coming from birth-death type of reactive events. One may also find it useful to further decompose the birth death noise $\sigma^{2}_{BD}$=$\sigma^{2}_{DMN+LNA}+\sigma^{2}_{corr}$ into a gaussian uncorrelated part $\sigma^{2}_{DMN+LNA}$ and higher orders of noise $\sigma^{2}_{corr}$.
 
\section{ Hybrid multi-scale stochastic simulation of genetic circuits }

The set of equations \ref{dmn} and \ref{dmn_lna} offer a multi-scale scheme for simulating genetic networks, where one accounts for the dichotomous gene noise at the master equation level, while generating the birth death noise either via deterministic dynamics in between stochastic gene state changes (DMN) or via a linear noise level of approximation (DMN+LNA).  There are many ways of simulating a set of piece-wise deterministic equations and here we use a method similar to the one proposed in the work of Alfonsi et al~\cite{alfonsi2004exact}. The state of a gene circuit $({\bf n},{\bf s})$ is represented by vectors of continuous ($\bf n$) and discrete species ({\bf s}). Since the deterministic protein evolution is coupled to stochastic gene switching, the rate coefficients ($k^s_{ij}(n(t))$) of stochastic switching ($s_i\rightarrow s_j$) become time dependent and one can not use a conventional Gillespie type Monte Carlo algorithm. The numerical procedure for simulating hybrid stochastic schemes (DMN/DMN+LNA) outlined in the section II consists of the following steps. First, one specifies the initial conditions for protein and gene states $({\bf n_0},{\bf s_0})$ for $t=0$. Then the dynamics is evolved by solving either an ODE (Eq~\ref{dmn}a ) or SDE (Eq ~\ref{dmn_lna}a) sets with a reasonably small time step $\Delta t$ which needs to be comparable or smaller than the typical time for gene switching $\tau_d=(k_{on}+k_{off})^{-1}$. All the simulations in the paper are carried out with $\Delta t=0.1\tau_d$. Given the initial conditions and the specified time step the trajectories are generated by numerically integrating the ODEs  or SDEs for each time $t_i=i*\Delta t_i$ with $i=[0,1,2...]$. The numerical integration continues until the condition for the stochastic gene switching ($s_i\rightarrow$ $s_j$) is reached. The switching condition is obtained by noting that the waiting time probability distribution for stochastic switching is given via $P(\tau)=exp\big(-\int^{t+\tau}_{t}\alpha (n,t)dt\big)$ where $\alpha=\sum_{j(\neq i)}k^s_{ij}$ is the escape rate from the state (${\bf n_i}$,$s_i$) and $\tau$ is the elapsed time after which stochastic event occurs. One can sample the stochastic transition events by generating a uniform random number $r_1$ on the interval $[0,1]$ and setting the condition for the stochastic switching as $r_1= exp\big(-\int^{t+\tau}_{t}\alpha (n,t)dt\big)$. If one has multiple genes in the system the criteria for choosing which one happens at the switching time is determined  by drawing a second random uniform number $r_2$ and testing for the condition $\sum^{i}_{j=1} k^s_{ij} \leq \alpha \cdot r_2 < \sum^{i+1}_{j=1} k^s_{ij}$ which determines the $s_i\rightarrow s_j$ gene switching event.

\subsection{Self regulating gene}

In this section we study the simplest model of the self regulating gene in order to explore and understand the ways white and dichotomous noise sources contribute to the overall noise level and also to test the quality of DMN and DMN+LNA approximations. The self-regulating gene circuit involves the following reactions:
\begin{subequations}
\begin{align}
ON \xrightarrow{k_{off}}  OFF+P \\
OFF+P \xrightarrow{k_{on}}  ON \\
P \xrightarrow{k} \emptyset \\
\emptyset \xrightarrow{g_{on}} P \\
\emptyset \xrightarrow{g_{off}} P
\end{align}
\end{subequations}
The first two reactions represent the gene switching processes which are responsible for the dichotomous noise, while the other reactions correspond to birth and death events.  The deterministic model for birth/death dynamics is $\dot n =g(s) -kn$, where $s$ stands for the gene state. The deterministic steady state attractors lead to differing protein numbers $n=\frac{g(s)}{k}$ for each gene state $s$. The linear noise approximation around each attractor is $dn = f(s,n)dt+\frac{g(s)}{k} dW$ where $dW$ stands for the Wiener process.  Varying the off rate $k_{off}$ while maintaining the relationship $k_{on}=0.1 k_{off}$ changes the adiabaticity with $\tau_d$ of the circuit smoothly going from the adiabatic to the non-adiabatic regime while not affecting the "equilibrium" properties of the binding process.  To quantify the protein noise level we compute the Fano factor $\sigma^2/\mu$. The Fano factor quantifies the deviation from Poissonian stochastic kinetics which corresponds to having a steady state distribution of protein numbers with $\sigma^2/\mu=1$.
\begin{figure}[!th]
\centerline{\includegraphics[width=.4\textwidth]{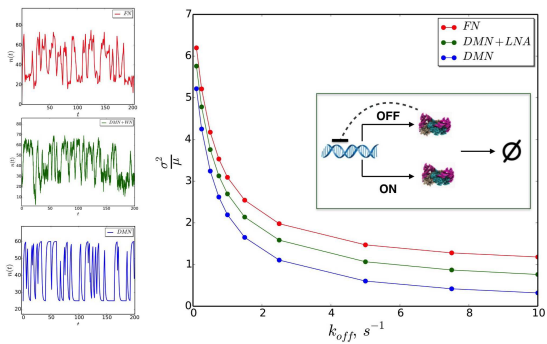}}
\caption{ Dependence of Fano factor on the unbinding rate of promoter ($k_{off}$) computed with dichotomous noise (DMN), DMN with linear noise (DMN+LNA) simulations and full noise simulations (FN) obtained by solving master equation. Inset shows the schematic topology of the model of self-regulated gene. Sample trajectories are shown on the left side. The values of the remaining rate coefficients are $g_{on}={25}$, $g_{off}=60$ and $k=1$.}
\label{fig:ff}
\end{figure}
The computed Fano factors shown in Fig. \ref{fig:ff} indicate that both the DMN and the DMN+LNA schemes do quite well in capturing the order of magnitude of the noise in both the adiabatic and non-adiabatic limits. This is quite remarkable, since the noise in DMN comes entirely from the local steady nonfluctuating states. This level of approximation does not account for the smallness of protein populations. Nevertheless the approximation provides a good  estimation of the observable Fano factors. The noise decomposition scheme outlined at the end of section II shows how noise from protein fluctuations caused by pure birth/death events compares to the total noise produced by the gene circuit as a whole: $\sigma^{2}_{BD}/\sigma^{2}_{tot}\sim 16 \%$ in the non-adiabatic regime ($k_{off}=0.1 s^{-1}$)  and  only $\sigma^{2}_{BD}/\sigma^{2}_{tot}\sim 70 \%$ even in the adiabatic regime ($k_{off}=10 s^{-1}$) for the present model circuit. Birth/death fluctuations become more important as the circuit approaches the adiabatic regime. Nevertheless we see that the contribution of dichotomous noise remains significant throughout this wide range of kinetic regimes. 
\begin{figure}[!th]
\centerline{\includegraphics[width=.4\textwidth]{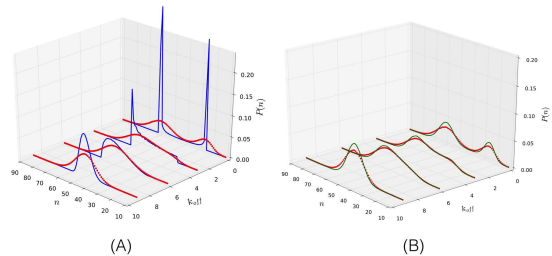}}
\caption{ Comparison of probability distributions obtained with full master equation (red lines) with those of  (A) DMN and (B) DMN+LNA  schemes for different values of adiabaticity parameter $k_{off}$.}
\label{fig:f3}
\end{figure}

Inspection of the detailed features of steady state probability distributions on the other hand reveals that pure dichotomous noise captures the skeleton of the distribution (Fig 3A), properly predicting whether the system's distribution is bimodal or unimodal along with even the relative weight of attractor states. Nevertheless  birth/death induced fluctuations in each steady state attractor are needed to give the distribution more reasonable widths. The peak heights are overestimated in the DMN approximation since the system does not fluctuate much in protein number around the steady state attractors ($g_0/k$ and $g_1/k$). The DMN+LNA scheme on the other hand does a much better job at reproducing the distribution with higher accuracy.  In the adiabatic limit both the DMN and the DMN+LNA schemes do quite well, although pure dichotomous noise alone naturally leads to more narrowly peaked distributions.
\begin{figure}[!th]
\centerline{\includegraphics[width=.4\textwidth]{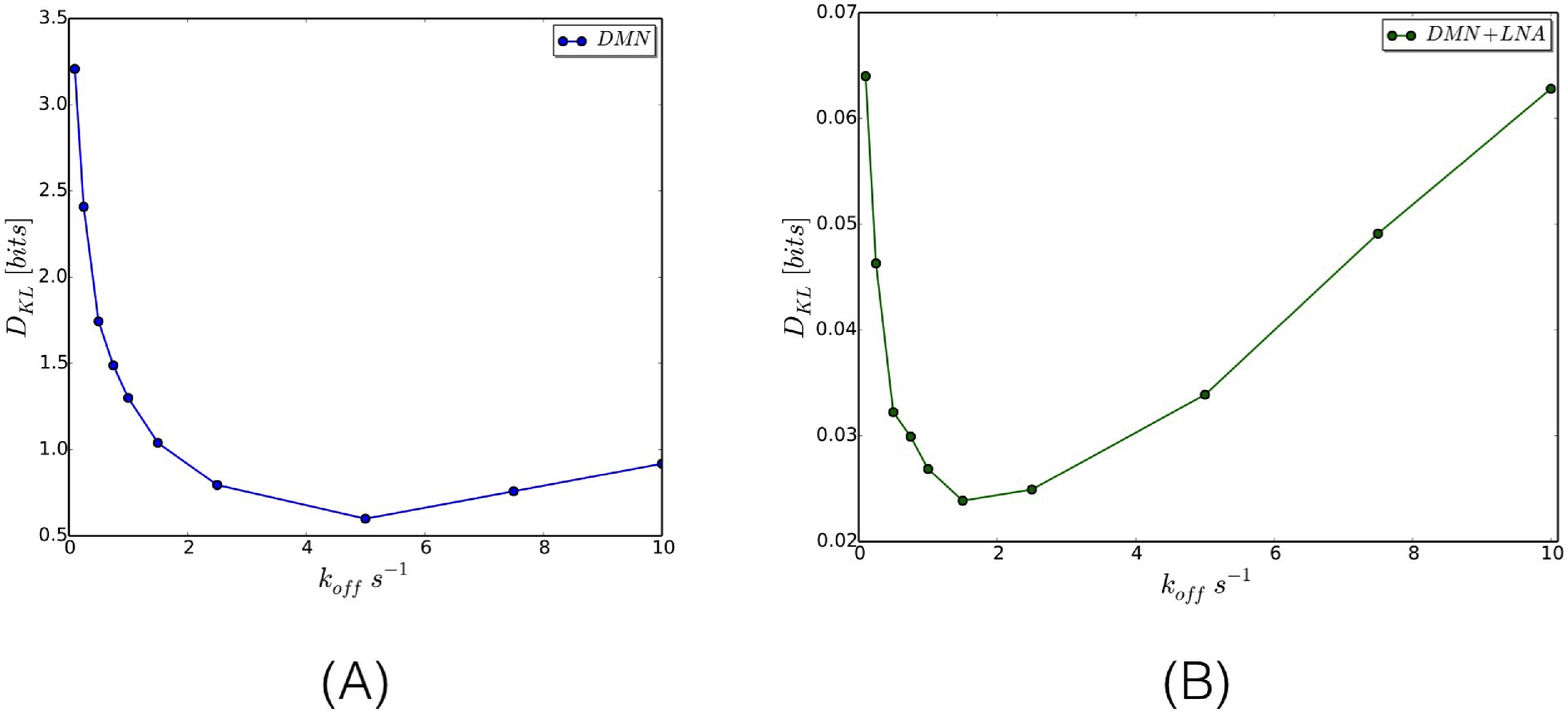}}
\caption{ The Kullback-Leibler divergence $D_{KL}(p | q)$ between exact distribution $p$ and approximate estimations $q$  for steady state probabilities obtained via (A) DMN and (B)  DMN+LNA simulations plotted against the adiabaticity parameter $k_{off}$. }
\label{fig:kl}
\end{figure}
A more stringent comparison between probability distributions is provided by the Kullback-Leibler (KL) divergence $D_{KL}(p | q)=\sum_x p(x)log\frac{p(x)}{q(x)}$, also known as the relative entropy in information theory~\cite{cover2012elements}.  $D_{KL}$ is a measure of statistical distinguishability between two probability distributions $p(x)$ and $q(x)$. It is a non-symmetric and positive function which vanishes only when the distributions are identical $p(x)=q(x)$. When $p(x)$ is the exact distribution directly obtained from the master equation and $q(x)$ is the approximate estimation via $DMN+LNA$ or $DMN$ schemes $D_{KL}(p|q)$ quantifies the amount of information in bits (see Eq.~\ref{fig:kl}) which is irrevocably lost due to approximations. From the comparison (Fig.~\ref{fig:kl}) we see that DMN+LNA is a better approximation in the intermediate range of gene switching and becomes worse in the very extreme limits. Thus the intermediate regime of gene switching is where such schemes are expected to shine. Many biochemical genetic circuits are expected to operate in this regime.

Overall these results are encouraging in that they show that one can obtain the steady state landscapes of large scale gene networks by only including dichotomous gene noise while modeling birth death events with the simple additive white noise. Simulating every reaction at a single molecule level is not only computationally prohibitive but probably also unnecessary for most cases. Purely deterministic models by themselves however are too crude for modeling gene networks. For instance the macroscopic model of a self regulating gene with  monomeric binding does not predict the possibility of bistable behavior in the non-adiabatic regime~\cite{grima2012steady, qian2009stochastic}, hence including gene states explicitly is crucial for exploring steady states of genetic circuits. 

\subsection{Extended toggle switch model and the importance of extinction states.}

In this section we consider a more complex and realistic circuit, the toggle switch. Toggle switches are ubiquitous bistable control modules. They have been implicated in regulating major cell fate decisions such as differentiation, cell death and development.  Some well known examples of genetic toggle switches include $\lambda$ phage's lysis/lysogeny switch, cell-cycle control circuits, signal transduction and stem cell differentiation pathways, etc. Here we will look closely at an extended toggle switch circuit model proposed by Strasser~\cite{strasser2012stability,marr2012multi}, which displays interesting multi-attractor dynamics. The circuit of the extended toggle switch consists of two genes A and B. It is based on a two state gene expression model with explicit transcription (producing mRNAs $M_A$ and $M_B$) and translation (producing  $P_A$ and $P_B$) steps. The proteins $P_A$ and $P_B$ are mutually inhibiting which gives rise to a bistable behavior. The reaction scheme of the extended toggle switch circuit contains the following elementary steps:  
\begin{subequations}
\begin{align}
ON_A \xrightarrow{k_{off_A}}  OFF_A+P_B \\
OFF_A+P_B \xrightarrow{kon_A}  ON_A \\
ON_B \xrightarrow{k_{off_B}}  OFF_B+P_A \\
OFF_B+P_A \xrightarrow{kon_B}  ON_B \\
OFF_A \xrightarrow{\beta_A} OFF_A+mRNA_A \\
OFF_B \xrightarrow{\beta_B} OFF_B +mRNA_B \\
mRNA_A \xrightarrow{g_{A}} mRNA_A+P_A \\
mRNA_B \xrightarrow{g_{B}} mRNA_B+P_B \\
mRNA_A \xrightarrow{\gamma_A} \emptyset \\
mRNA_B \xrightarrow{\gamma_B} \emptyset \\
P \xrightarrow{\delta_A} \emptyset \\
P \xrightarrow{\delta_B} \emptyset \\
\end{align}
\end{subequations}
Reactions (a-d) are gene switching events,  (e,f) correspond to transcription process, (g,h) to translation process and the remaining reactions describe the degradation of mRNA and protein molecules. The deterministic equations describing the birth-death processes are:
\begin{widetext}
\begin{subequations}
\begin{align}
\dot M_A =f_1({\it s_A},M_A)= {\it s_A}  \beta_A - \gamma_A M_A \\
\dot M_B = f_2({\it s_B},M_B)={\it s_B}  \beta_B - \gamma_B M_B\\ 
\dot P_A = f_3({\it s_B},M_A,P_A)=g_A M_A-\delta_A  P_A +k_{off_B} {\it s_B}-k_{on_B}  (1-{\it s_B})  P_A \\
\dot P_B =f_4({\it s_A},M_B,P_B)=g_B M_B-\delta_B  P_B +k_{off_A}  {\it s_A} -k_{on_A}  (1-{\it s_A})  P_B 
\end{align}
\end{subequations}
\end{widetext}
We study the case of equal rate coefficients for  genes $A$ and $B$, making the toggle switch symmetric. 
\begin{subequations}
\begin{align}
M^{ss}(s)=\frac{\beta  \cdot {\it s}}{\gamma} \\
P^{ss}(s)= \frac{g  M^{ss}(s') + k_{off}  (1-{\it s})}{\delta+k_{on} \cdot {\it s}}
\end{align}
\end{subequations}
The variable $s$ stands for the gene state of the first gene (0 for $OFF$ and 1 for $ON$ ) and ${s'}$ denotes the state of the second gene.  The DMN+LNA equations are:
\begin{widetext}
\begin{subequations}
\begin{align}
dM_A(t) = f_1({\it s_A},M_A)dt + {\it s_A} \sqrt{\beta_A} dW_1- \sqrt{\gamma_A \cdot M^{ss}({\it s_A})} dW_2 \\
d M_B(t) = f_2({\it s_B},M_B)dt +{\it s_B} \sqrt{\beta_B} dW_3- \sqrt{\gamma_B \cdot M^{ss}({\it s_B})} dW_4 \\ 
dP_A(t) =  f_3({\it s_B},M_A,P_A)dt+\sqrt{g_A M^{ss}({\it s_A})}dW_5-\sqrt{\delta_A \cdot P^{ss}(s_B)}dW_6 \\
dP_B(t) =f_4({\it s_A},M_B,P_B)dt+  \sqrt{g_B M^{ss}({\it s_B})}dW_7-\sqrt{\delta_B \cdot P^{ss}(s_A)}dW_8
\end{align}
\end{subequations}
\end{widetext}
The stochastic transitions of the gene state {\it s} are again governed by a master equation.
\begin{figure}[!h]
\centerline{\includegraphics[width=.4\textwidth]{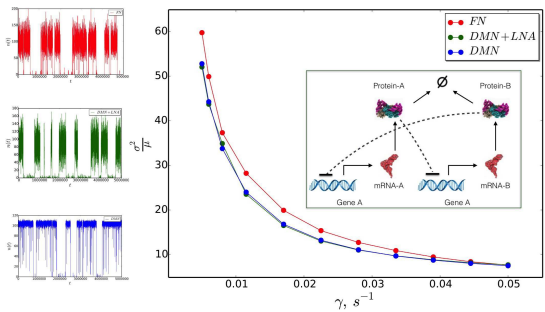}}
\caption{ Dependence of Fano factor on the mRNA degradation rate $\gamma$ computed with full noise (FN) dichotomous noise(DMN) and DMN with linear noise correction (DMN+LNA). Inset shows the schematic topology of the model of extended toggle switch. Sample trajectories are shown on the left side. The values of the remaining rate coefficients are $g_{on}={0.05}$, $g_{off}=0$, $\delta=0.005$, $k_{on}=5$, $k_{off}=0.1$ and $\beta=0.05$ for both A and B genes. }
\label{fig:toggleff}
\end{figure}
In the present model the basal rate of protein production in the repressed bound state has been set to zero ($g_{off}=0$), which creates strongly bistable behavior between extinction and steady production. The extinction attractors correspond to  small or virtually non-existent mRNA and protein populations.  Such extinction states are interesting to study with our methods both for understanding the how different types of noise sources impact the stability of bistable switches and also for finding the true limitations of our approximation schemes. Both the DMN and DMN+LNA equations are based on deterministic evolution of molecular populations and they either completely neglect the birth/death fluctuations or treat them approximately.  It is interesting to see how such schemes would behave in the regimes where the birth/death fluctuations are significant. By varying the mRNA degradation rate $\gamma$ we vary the average lifetime of mRNAs which controls the burst size of proteins. Thus we go from the regime of high birth/death noise (small $\gamma$ and $\sim10$ molecules) to low birth/death noise (large $\gamma$ and $\sim10^2$ molecules). 
\begin{figure}[!th]
\centerline{\includegraphics[width=.4\textwidth]{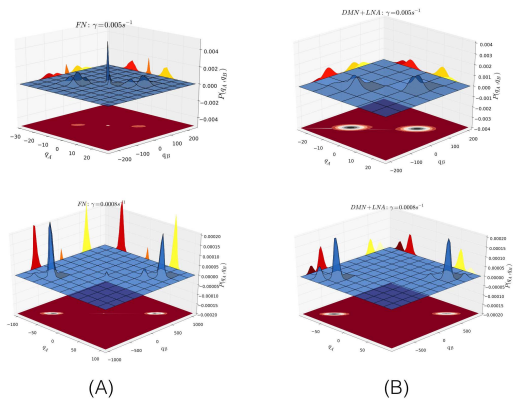}}
\caption{ Steady state probability distribution projected in the space of $q_A=N(mRNA_A)-N(mRNA_B)$ and $q_B=N(P_A)-N(P_B)$ order parameters reveals the multi-attractor nature of extended toggle switch. Comparing (A) Full Noisewith (B) DMN+LNA approximations for the different values of mRNA lifetime $\gamma$.}
\label{fig:toggleproj}
\end{figure}
The dependence of the Fano factor on $\gamma$ shows (Fig.~\ref{fig:toggleff}) that indeed both the DMN and DMN+LNA schemes capture the total noise levels of the circuit quite well with the agreement improving steadily with the decreasing contribution of the birth/death noise. Detailed inspection of the probability distributions however reveals that the DMN+LNA scheme fails to capture metastable near-extinction states. To better visualize all of the steady state attractors we have projected the 2D probability distribution on the the order parameters $q_A=M_A-M_B$ and $q_B = P_A-P_B$ which emphasize the importance of states with $M_A \approx M_B \sim10$ and $P_A \approx P_B\sim 10$,  where  both proteins and mRNAs are present in equally low quantities. From the probability distribution at high birth/death noise limit obtained from the simulation with full noise(see the top figure in Fig.~\ref{fig:toggleproj}A)  we see that the toggle switch displays multi-attractor behavior with two main attractors corresponding to two configurations of the toggle switch with gene states  $s_A=ON(OFF)$ and $s_B=OFF(ON)$ and the metastable attractor corresponding to both genes being simultaneously repressed ($s_A=s_B=ON$) which gives rise to a narrow peak in the center of 2D probability distribution. 

The steady state probability distributions obtained with the DMN+LNA scheme capture the main attractors in the system but do not reveal the presence of metastable states. This is because in the linear nose approximation of the birth death noise the fluctuations of discrete species are treated as uncorrelated brownian noise terms. Thus the correlated fluctuations that are obtained by treating the noise at the full master equation level are essential to the existence of the extinction state attractors in this system. Such metastable attractors while conceptually interesting have a negligible contribution to the total noise level. They are only observed when the proteins and mRNAs are present in extremely low quantities on the order of $\sim10$. Once the numbers of molecules reach levels of $\sim100$ the DMN+LNA scheme provides an excellent approximation to the full master equation as can be seen from the comparison of distributions on the lower panel of Fig.~\ref{fig:toggleproj}. If one is interested in studying such metastable states with hybrid stochastic techniques, one possibility would be to treat the reactions involving extremely small numbers molecules into the space of discrete variables and treating them at the same master equation level as the gene states.

\begin{figure}[!th]
\centerline{\includegraphics[width=.4\textwidth]{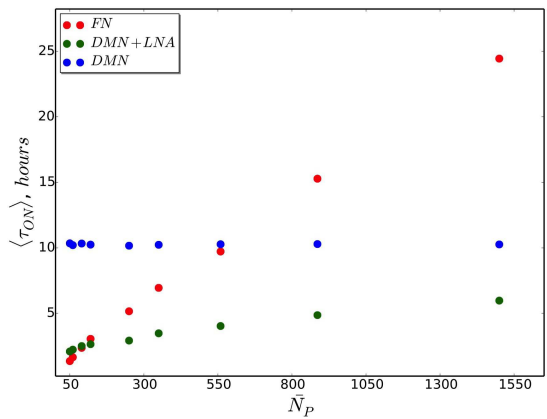}}
\caption{Residence times in the $ON$ state of a gene vs the protein population size of an extended toggle switch model.}
\label{fig:togglesize}
\end{figure}

Finally we study how the  molecular population size affects the residences times of a gene being in the ON state which provides a measure of stability of an attractor. We vary the  population size by changing the translation rate ($g=g_A=g_B$) (Fig.~\ref{fig:togglesize}) for both genes. In the case of the full noise the residence times grow linearly with the population size (number of Proteins and mRNAs)  which is in agreement with the previous studies on the extended toggle switch circuit~\cite{strasser2012stability,marr2012multi}. Overall due to the white noise contribution the noise at the DMN+LNA level allows for more fluctuations in the gene state which in turn results in more transition events. 
This aspect allows faster exploration of gene circuits distribution, a feature that may be useful for quick scanning for the steady state attractors in large genetic networks. The residence times computed with the DMN  scheme are obviously insensitive to the changes of protein numbers. This insensitivity is due to the  deterministic treatment of these reactions involving species other than genes which do not  affect the stochastic gene switching events.

\subsection{Conclusion}
In this work we have explored hybrid stochastic schemes based on treating the genetic noise of DNA binding explicitly while making  linear noise approximations for the birth and death reactions of other species. The theoretical motivation for such schemes along with the key assumptions and connections of hybrid schemes with the underlying master equation are elucidated in a path sum based analysis of trajectory statistics. We have explored the steady state attractors of two classic toy models of gene circuits, the self-regulating gene and an extended version of a genetic toggle switch. Our findings indicate that gene state change noise accounts for a large fraction of the total noise for a wide range of time-scales. The noise decomposition obtained by comparing the total noise estimated by the different schemes for the extended model of toggle switch circuit revealed that the noise induced metastable attractor originates from correlations of discrete molecular populations. Our study suggests that stochastic hybrid schemes will be useful tools for exploring large scale genetic networks, which display complex multi-attractor dynamics caused by discrete fluctuations in the promoter architecture and should find their place alongside more traditional full scale kinetic Monte Carlo simulations. For future work it would be interesting to use such hybrid schemes to model the effects of gene co-localization, heterogeneous  distribution of transcription factors in nuclear/cytoplasmic milieu and 1D/3D diffusional motions by adding additional deterministic steps for each molecular component and enlarging the number of internal discrete promoter states. The latest studies have shown that in eukaryotic cells gene expression is an intricate function of chromatin configuration as well as spatial distribution of decoy binding sties on the genome. Thus it would be exciting to use hybrid schemes accounting for internal gene states with a high level of detail via a discrete web of Markov states to get a good understanding of some of these complex facets in eukaryotic gene regulation.  At last a possible extension to hybrid schemes with coupled diffusion introduced for selected species could also be an attractive way of treating in a coarse grained manner the cases where the well stirred assumption does not hold. 



\section*{Acknowledgments}

We gratefully acknowledge financial support by the D.R. Bullard-Welch Chair at Rice University, Grant C-0016 and PPG Grant P01 GM071862 from the National Institute of General Medical Sciences.



%
%

%


\bibliography{ref}

\end{document}